# Electron Paramagnetic Resonance Study of Radical Species on $NaNbO_3$@$CeO_2$-Modified Carbon Vulcan XC72 Gas Diffusion Electrode for Electrochemical Degradation of Paracetamol via Electro-Fenton

Caio Machado Fernandes[1], João Paulo C. Moura[1], Aline B. Trench[1], Rafael Sotana[2], Ana Maria P. Neto[2], Willy G. Santos[1], Mauro C. Santos[1*]

[1]*Centro de Ciências Naturais e Humanas, Universidade Federal do ABC – Santo André, SP – Brasil*

[2]*Centro de Engenharia, Modelagem e Ciências Sociais Aplicadas, Universidade Federal do ABC – Santo André, SP – Brasil*

Corresponding Author:

*E-mail: mauro.santos@ufabc.edu.br (M.C. Santos).

**Abstract**

While electrochemical oxidation is a promising technology for water treatment, a fundamental understanding of the specific radical mechanisms involved in pharmaceutical degradation has remained limited. This study addresses this gap by employing Electron Paramagnetic Resonance (EPR) spectroscopy to directly quantify the radical species generated during the degradation of paracetamol using a novel gas diffusion electrode (GDE) modified with $NaNbO_3$ nanocubes and $CeO_2$ nanorods. This approach provides a critical advancement beyond prior literature by moving from indirect inference to direct, quantitative analysis of reactive species. Results demonstrated that a boron-doped diamond (BDD) anode (65% $^{\bullet}OH$, 35% aryl radicals) drastically outperformed a Platinum (Pt) anode (74% $^{\bullet}OH$, 26% aryl radicals), achieving complete degradation in 15 minutes versus 45 minutes and 81.6% versus 67.8% mineralization. Consequently, this work provides a foundational mechanistic framework that fundamentally advances the field, offering not just a more effective material system (BDD/$NaNbO_3$@$CeO_2$-GDE) but also a validated methodology for rationally designing and optimizing electrochemical water treatment processes based on quantifiable radical pathways.

## 1. Introduction

N-acetyl-p-aminophenol, pharmaceutically known as acetaminophen or Paracetamol, is extensively used for analgesic and antipyretic purposes. The extensive use of this substance leads to significant environmental contamination due to human excretion, improper disposal practices, and effluents from wastewater treatment plants. This widespread release poses serious risks to our ecosystems and underscores the urgent need for effective management strategies [1].

Recent studies have shown that chronic exposure to paracetamol can affect liver and kidney function in wildlife and, potentially, in humans. Furthermore, paracetamol ion may interact with some minerals, elements, and/or organic compounds in water, leading to complex adsorption behaviors that complicate removal. These interactions can enhance its persistence and bioavailability in the environment, necessitating the development of advanced treatment methods to mitigate its ecological and health impacts [2,3].

Conventional techniques for removing paracetamol from water media, such as biological treatment and other adsorption processes, have proven insufficient for health purposes due to the compound's persistence and resistance to degradation. These methods often fail to eliminate paracetamol, leading to its accumulation in aquatic environments and posing ongoing ecological and health risks. Electrochemical advanced oxidation processes (EAOPs) have emerged as an up-and-coming solution for efficiently removing persistent organic compounds, including paracetamol [4-8].

$$O_2 + 2H^+ + 2e^- \rightarrow H_2O_2 \quad (1)$$

$$H_2O_2 + Fe^{2+} \rightarrow Fe^{3+} + {}^{\cdot}OH + {}^{-}OH \quad (2)$$

$$Fe^{3+} + e^- \rightarrow Fe^{2+} \qquad E^o = -\,0.77\ V/SHE \quad (3)$$

The performance of the electro-chemical method is fundamentally influenced by the nature of the elements involved and the chemical properties of the catalyst used, which plays a crucial role in enhancing catalysis efficiency. An effective catalyst significantly improves the generation of $H_2O_2$ molecules by optimizing the Oxygen Reduction Reaction (ORR) [13]. This optimization results in higher yields of electrogenerated $H_2O_2$ amount, high values for reaction rates, and greater chemical stability, ensuring a consistent and effective degradation process [14,15]. Additionally, an efficient catalyst helps to reduce energy consumption and operational costs, making the electron-Fenton method more economically and environmentally sustainable for large-scale applications.[16,17].

Cerium and niobium ions have emerged as promising electrocatalysts for the two-electron oxygen reduction reaction (ORR) due to their low potential values, which are crucial for the in situ electro-generation of hydrogen peroxide ($H_2O_2$) in the electro-Fenton process. Cerium, known for its excellent redox properties, can undergo reversible $Ce^{3+}/Ce^{4+}$ transitions. Its ability to create binding sites with oxygenated species (*e.g.,* OR) such as $H_2O$ or $OH^-$ anion facilitates efficient electron transfer processes in the electrode, thereby improving hydrogen peroxide generation. The niobium ($Nb^{3+}$) ion has a favorable electron arrangement to form an active and stable surface for the selective two-electron ORR pathway, thereby increasing the catalytic response for $H_2O_2$ generation [18-22].

Building on the foundation of single-metal catalysts, recent research has increasingly focused on sophisticated nanocomposites and advanced diagnostic techniques to further enhance the electro-Fenton process. For instance, studies exploring novel heterostructures have demonstrated how interfacial engineering can optimize charge separation and create synergistic active sites, leading to superior $H_2O_2$ production and pollutant degradation [23-25]. A critical challenge in leveraging these advanced materials lies in precisely quantifying the radical species responsible for degradation, as this determines the underlying mechanism and efficiency. Traditional methods often infer radical presence indirectly; however, the application of Electron Paramagnetic Resonance (EPR) spectroscopy has emerged as a powerful tool for the direct identification and semi-quantification of reactive oxygen species, providing unambiguous evidence that advances the field beyond speculation.

Considering all the issues related to before, this work aims to enhance the understanding of paracetamol degradation by investigating the effectiveness of a novel gas diffusion electrode made from carbon Vulcan XC72 modified with $NaNbO_3$@$CeO_2$. By harnessing the principles of electrochemical advanced oxidation processes (EAOPs), particularly the electro-Fenton method, the research seeks to improve the degradation efficiency of paracetamol. The choice of $NaNbO_3$ and $CeO_2$ as unique electrocatalyst materials is grounded in their favorable redox properties, which were used to promote the oxidation process by generation of reactive oxygen species. Furthermore, this study employs the technique of electron paramagnetic resonance (EPR) to identify the main radical species involved in the degradation of paracetamol, demonstrating the dynamic of radicals during the electro-Fenton reaction.

## 2. Experimental and Methods

### 2.1. Synthesis of the $NaNbO_3/CeO_2$ based nanomaterial

The $NaNbO_3$ microcubes were prepared and decorated with $CeO_2$ nanorods according to a previous work published in our research group [26]. 1 g of $Nb_2O_5$ and 12 g of NaOH were dispersed in 30 mL of water and stirred for 30 minutes at room temperature (~ 300 K). The mixture was then transferred to a stainless-steel reactor, which was sealed and heated at 523 K for 72 hours. Once cooled to room temperature, the resulting residue was three times centrifuged in 0.1 mol $L^{-1}$ HCl solution. The solid was subsequently washed with water and centrifuged repeatedly to reach pH values around 7.0. The final solid product was dried at 373 K for 24 hours. The microscopy images of the synthetized material were previously reported and discussed by Antonin and co-workers [26].

### 2.2. Gas diffusion electrode preparation

The gas diffusion electrode (GDE) preparation involved the hot-pressing method. The catalytic mass comprised carbon Vulcan XC72 modified with 1% (w/w) $NaNbO_3@CeO_2$, and polytetrafluoroethylene (PTFE). PTFE is a hydrophobic binder in 20% related to electrocatalytic material. The concentration of 1% chosen for the nanomaterial was according to the results obtained previously and published in the literature [26]. Cylindrical GDEs with a diameter of 25 mm were fabricated using the following procedure: initially, a stainless-steel plate was positioned at the base of the mold. Subsequently, 2.0 g of the prepared electrocatalyst was introduced. Finally, another stainless steel plate was placed at the top of the catalyst layer. The assembly was subjected to compression under a load of 2.5 tons for two hours while maintaining the mold temperature at 563 K [27].

### 2.3. Paracetamol degradation

#### 2.3.1 Electrochemical reaction and UV-Vis measurements

A polypropylene undivided cell was built especially for the GDE (working electrode) system and was used to perform the electro-Fenton experiments regarding the degradation of paracetamol. Free oxygen solution of 350 mL of $K_2SO_4$ (0.1 mol $L^{-1}$ and pH = 3), $H_2SO_4$ (0.5 mmol $L^{-1}$), and paracetamol (50 mg $L^{-1}$) were prepared and used in all electrochemical experiments. Platinum (Pt) or boron-doped diamond (BDD) was used as a counter electrode (anode). The electro-Fenton process was performed with the fixed potential of -2.7 V, using Ag|AgCl as reference electrode. In agreement with our previous work [26], the potential of -2.7V was chosen because of the high efficiency to

generate $H_2O_2$, using the same conditions of GDE and chemical media. Kinetics of degradation of Paracetamol at different times of electrolysis (0, 15, 30, 45 and 60 min) was investigated by UV-Vis spectroscopy, following the absorbance value of the lowest energetic band centered at 243 nm. Every measurement was done in triplicates.

### 2.3.2 Electron paramagnetic resonance (EPR) experiments

X-band (9.45 GHz) CW-EPR measurements were carried out using a Bruker EMX plus spectrometer at 298 K in a suprasil quartz tube with 100 kHz magnetic field modulation of 1.0 G amplitude, magnetic field sweep (100 G), modulation amplitude (1.0 G), microwave power between 1 and 15 mW (usually 5mW was fixed for few experiments), conversion time (40.96 ms), and receiver gain ($1.0 \times 10^5$) were used in all experiments and/or simulations.

In a typical experiment, the spin trap compound (DMPO, TMPO) was added ($10^{-4}$ mol $L^{-1}$) to the solution of 100 µL of the electrochemical reaction containing paracetamol (N-acetyl-para-aminophenol). The simulated spectra were performed, and the hyperfine constants were used to identify the radical species.

### 2.3.3 Total organic carbon (TOC) measurements

TOC measurements were performed using a TOC-L analyzer (Shimadzu, Kyoto, Japan). Samples were introduced into a combustion tube and heated to 680°C. The combustion process converted the organic carbon in the sample to carbon dioxide ($CO_2$). Synthetic air, at a 150 mL/min flow rate, carried the combustion products to an electronic dehumidifier for cooling and dehydration. The gas stream was then purified to remove chlorine and other halogens before being transported to a non-dispersive infrared (NDIR) sensor for $CO_2$ detection. The TOC-Control L software recorded the NDIR signal peak area, which is directly proportional to the carbon concentration in the sample.

The energy consumption (EC) and the mineralization current efficiency (MCE) were calculated according to eq. 4 and 5:

$$\mathrm{EC} = \frac{\mathrm{E_{cell}\, I\, t}}{\mathrm{V_s(TOC_o - TOC)}} \tag{4}$$

$$\mathrm{MCE\ (\%)} = \frac{n\, V_s\, F\ (\mathrm{TOC_o - TOC})}{4.32e^7\ m\, I\, t} \times 100 \tag{5}$$

Where $TOC_o$ is the initial total organic carbon, TOC is the total organic carbon at time t, n is the number of electrons involved in the complete mineralization of a paracetamol molecule (16), *F* is the Faraday constant (96487 C $mol^{-1}$), and m is the number of carbon atoms of paracetamol.

## 3. Results and discussion

### 3.1. Electro-Fenton degradation of paracetamol

The electro-Fenton degradation of paracetamol aqueous solution (50 mg $L^{-1}$) was investigated by use of GDE composed of Vulcan XC72 carbon modified with $NaNbO_3@CeO_2$ as a cathode. Pt or BDD was used as an anode electrode. The electrochemical experiment was carried out at pH 3.0 with the supporting electrolyte of 0.1 mol $L^{-1}$ $K_2SO_4$, in the presence of 0.5 mmol $L^{-1}$ $Fe^{2+}$ and potential value fixed at -2.7 V.

Fig. 1 shows the shows the normalized paracetamol concentration vs. time plots and the kinetic profiles for the degradation of the paracetamol compound using Pt and BDD electrodes. The concentration ratio of $[C]/[C_0]$ decreases exponentially with electrolysis time for both electrodes, which is characteristic of a first-order kinetic process as also described by the following equation:

$$ln\left(\frac{C}{C_0}\right) = -kt$$

From the linear fitting of the experimental data (inset in Fig. 1), the apparent rate constants (k) were determined. For the Pt electrode, k = 0.077 ± 0.002 $min^{-1}$, whereas for the BDD electrode, the rate constant was significantly higher, k = 0.345 ± 0.011 $min^{-1}$. These results clearly demonstrate that the BDD electrode exhibits a markedly superior electrocatalytic activity compared to Pt, leading to 4.48 times ($k_{BDD}/k_{Pt}$ = 4.48 ) faster degradation rate under identical experimental conditions. This enhanced performance can be attributed to the higher oxygen evolution overpotential and the more efficient generation of oxidizing species (e.g., hydroxyl radicals) on the BDD surface, which favor indirect oxidation pathways and improve the overall degradation efficiency.

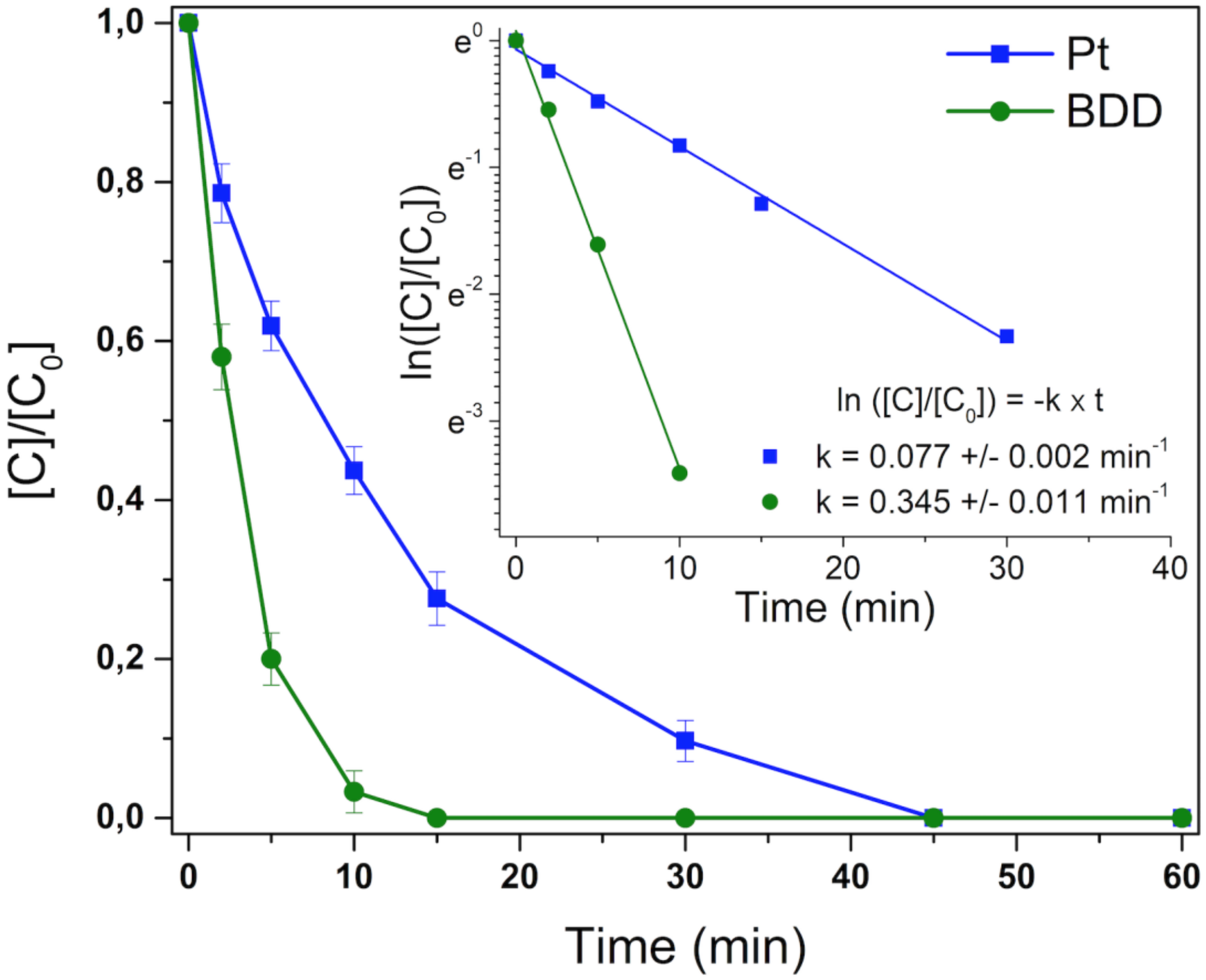

**Figure 1**. Normalized paracetamol removal vs. electrolysis time for the electro-Fenton process ($[C_0]$ = 50 mg $L^{-1}$, 0.1 mol $L^{-1}$ $K_2SO_4$ aqueous solution, pH = 3.0, E = -2.7 V, 0.5 mmol $L^{-1}$ $Fe^{2+}$) with Vulcan XC72 carbon modified with $NaNbO_3$@$CeO_2$ GDE as cathode and Pt or BDD as anode.

The kinetic trace of paracetamol degradation also indicates that approximately 55% of the paracetamol molecule is quickly degraded at the initial times (10 min) of electrolysis, when Pt is used as anode electrode, which suggests a sustained generation of reactive hydroxyl radicals ($^{\bullet}OH$) facilitated by the efficient catalytic activity of NaNbO3@CeO2-modified cathode [28]. At 30 minutes, the removal percentage reaches nearly 85%, with a slight deceleration in the degradation rate likely due to the reduced availability of paracetamol molecules. Complete removal of paracetamol is observed by 45 minutes, demonstrating the high efficiency of the electro-Fenton process under these conditions. The acidic environment, optimal for the Fenton reaction, and the enhanced surface area and catalytic properties of the modified GDE cathode likely contribute to the sustained production of $^{\bullet}OH$, which drives the degradation of paracetamol [29, 30].

When BDD is used as an anode electrode, 80% of paracetamol is fast degraded at over 5 min of reaction, indicating an exceptionally rapid degradation rate, fast kinetics associated with the generation of $^{\bullet}OH$ in bulk by electro-Fenton reaction and the BDD($^{\bullet}OH$). This trend continues with the complete removal of paracetamol achieved in 15 minutes, confirming the BDD anode's ability to maintain a highly reactive state that sustains the generation of radical species. Unlike platinum anodes, where $^{\bullet}OH$ is easily oxidized, the BDD anode effectively accumulates $^{\bullet}OH$ species on its surface, enabling their direct involvement in the oxidation of paracetamol molecules. This accumulation of $^{\bullet}OH$ radicals enhances the electro-Fenton process, increasing the molecular encounter between radical and paracetamol molecules by a diffusion process, leading to paracetamol's complete degradation and potential mineralization [31-33].

Compared to the results obtained with the Pt anode, a BDD anode significantly accelerates the process (total degradation in 15 minutes versus 45 minutes). The presence of BDD, known for its superior electrocatalytic properties and high oxygen evolution overpotential, contributes to a more rapid and effective generation of reactive species, particularly $^{\bullet}OH$ radicals, which are critical for the degradation of paracetamol. Additionally, BDD anodes allow for the direct oxidation of organic pollutants at the anode surface, further enhancing the degradation efficiency. This dual mechanism—combining indirect oxidation through $^{\bullet}OH$ and direct oxidation on the BDD surface—results in the much quicker attainment of complete removal with a BDD anode. While Pt anodes are adequate, the broader operational window, stronger oxidative environment, and direct oxidation capability provided by the BDD anode make it a more potent choice for applications requiring rapid and thorough degradation of pharmaceutical pollutants like paracetamol in wastewater treatment processes [34-38].

#### 3.1.1. Total Organic Carbon (TOC) abatement

The TOC analysis provides essential insights into the mineralization efficiency of the electro-Fenton process in the degradation of paracetamol. The TOC results are shown in Fig. 2a and reveal a progressive reduction in organic carbon content, indicating the extent to which organic molecules, including paracetamol and its degradation intermediates, are mineralized [39]. Fig. 2b shows the energy consumption, and Fig. 2c shows the mineralization current efficiency, both for the paracetamol degradation via electro-Fenton with a $NaNbO_3$@$CeO_2$ GDE.

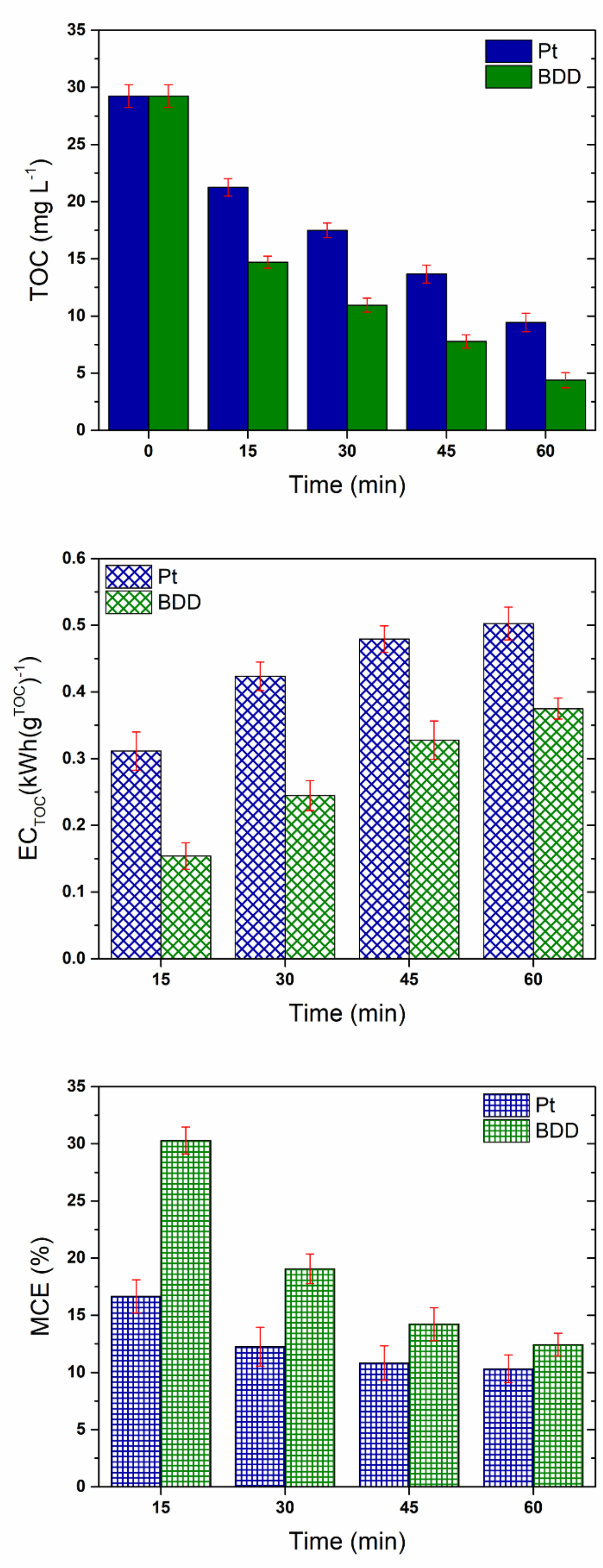

**Figure 2**. Total organic carbon vs. electrolysis time (a), energy consumption (b), and mineralization current efficiency (c) for the electro-Fenton process (Paracetamol [$C_0$] = 50 mg $L^{-1}$, 0.1 mol $L^{-1}$ $K_2SO_4$ aqueous solution, pH = 3.0, E = -2.7 V, 0.5 mmol $L^{-1}$ $Fe^{2+}$) with Vulcan XC72 carbon modified with $NaNbO_3$@$CeO_2$ GDE as a cathode and Pt or BDD as an anode.

Regarding Fig. 2a, when Pt was used as an anode, the TOC decreased from 29.24 to 21.25 mg $L^{-1}$ after 15 minutes of electrolysis, representing a 27.3% reduction in TOC. This reduction at the early stages of the process suggests that paracetamol undergoes rapid degradation, but the oxidation products include various intermediate organic compounds that contribute to the remaining TOC. The hydroxyl radicals generated through the electro-Fenton process effectively attack the parent paracetamol molecules and their intermediates, such as hydroquinone, benzoquinone, or smaller organic acids. However, the presence of carboxylic acids and other intermediates in the solution indicates that high mineralization values are not immediately achieved [40].

By 30 and 45 minutes, the TOC further decreased to 17.49 and 13.67 mg $L^{-1}$, reflecting a 40.2 and 53.2% reduction in TOC, respectively. This continued decline indicates that the intermediate products formed during the initial degradation phase undergo further oxidation. The Vulcan XC72 carbon modified with $NaNbO_3$@$CeO_2$ cathode plays a critical role in sustaining the production of $^{\bullet}OH$ radicals, which break down these intermediates into simpler organic molecules or mineralize them into $CO_2$. The slower rate of TOC reduction compared to the initial degradation of paracetamol suggests that while the main compound is quickly broken down, its byproducts require extended exposure to an oxidative environment for further mineralization [41, 42].

At 60 minutes, the TOC dropped to 9.44 mg $L^{-1}$, representing a 67.8% reduction in TOC. This significant reduction demonstrates the efficacy of the electro-Fenton process in degrading and mineralizing organic pollutants over time. By this stage, the intermediates generated during the initial breakdown of paracetamol have likely undergone further oxidation, leading to the formation of lower-molecular-weight compounds or complete mineralization into $CO_2$ and water. However, the rate of TOC reduction slows as the concentration of easily oxidizable organics diminishes and the remaining compounds become more recalcitrant or near complete mineralization [43].

When BDD was used as an anode, the TOC decreased from 29.24 to 14.71 mg $L^{-1}$ after 15 minutes, indicating a 49.7% reduction in TOC. This substantial reduction at an early stage highlights the BDD anode’s ability to help degrade paracetamol and effectively mineralize its organic intermediates. The high efficiency is attributed to the strong oxidative environment provided by the BDD anode, which supports both the generation of hydroxyl radicals and the direct oxidation of organic pollutants at the anode surface. This dual mechanism significantly enhances the mineralization process, reducing the presence of intermediate organic compounds in the solution[44, 45].

As the electrolysis progressed to 30 and 45 minutes, the TOC further decreased to 10.95 and 7.77 mg $L^{-1}$, representing a 62.6 and 73.4% reduction in TOC, respectively. This continued decline indicates that the BDD anode, alongside the Vulcan XC72 carbon modified with $NaNbO_3$@$CeO_2$ GDE cathode, maintains a highly oxidative environment that sustains the oxidation of organic intermediates and facilitates their further breakdown into simpler compounds. This leads to a more efficient mineralization pathway, significantly reducing the total organic carbon content over time [46].

By 60 minutes, the TOC dropped to 4.38 mg $L^{-1}$, indicating an 85.0% reduction in TOC. This substantial reduction reflects the BDD anode's ability to achieve near-complete mineralization of paracetamol and its degradation products. The combination of $^{\bullet}OH$ production and direct oxidation at the BDD surface ensures that even the more recalcitrant organic intermediates are effectively broken down, leading to their eventual mineralization. The steady decrease in TOC throughout the experiment underscores the high efficiency of the BDD anode with the $NaNbO_3$@$CeO_2$ GDE in promoting both degradation and mineralization of organic pollutants in the electro-Fenton process [47-49].

When comparing these results with those obtained with a Pt anode, the superiority of the BDD anode becomes evident. The BDD not only accelerates the degradation of paracetamol but also significantly enhances the mineralization of the resulting intermediates. This increased efficiency can be attributed to the BDD anode's unique properties, high oxygen evolution overpotential, and strong oxidative capabilities. Unlike Pt anodes, which rely primarily on generating $^{\bullet}OH$ radicals for the oxidation of organics, BDD anodes allow for direct oxidation of organic pollutants at the anode surface. This direct oxidation mechanism is particularly effective in breaking down more complex and recalcitrant organic molecules, leading to faster and more complete mineralization. Consequently, the BDD anode facilitates anodic oxidation, accelerating the treatment process when combined with the electro-Fenton method. This reduces the time required for significant TOC reduction and ensures a higher overall level of mineralization, making it a more practical option for treating pharmaceutical pollutants like paracetamol in wastewater [50-53].

The TOC data highlights that while paracetamol is rapidly degraded, complete mineralization is more gradual. Initial degradation primarily involves breaking the aromatic ring and other molecular structures, forming various organic intermediates. The effectiveness of the Vulcan XC72 carbon modified with $NaNbO_3$@$CeO_2$ cathode in producing $^{\bullet}OH$ radicals from electrogenerating $H_2O_2$ and allowing the electro-Fenton process to occur is crucial throughout the process, particularly in sustaining the oxidation of these intermediates. However, the incomplete TOC removal after 60 minutes indicates that some intermediates remain resistant to oxidation or require more prolonged exposure to fully mineralize, underscoring the complexity of organic pollutant degradation in electrochemical processes.

The energy consumption (EC) and mineralization current efficiency (MCE) values, calculated from the TOC data, are depicted in Fig. 2b and 2c, respectively. In both cases, the MCE values decreased over time, primarily due to two factors: the progressive loss of organic matter available for oxidation and the formation of more recalcitrant by-products that are harder to mineralize. As the electrolysis progresses, the remaining organic intermediates become increasingly resistant to oxidation, requiring more energy and time for complete mineralization. This trend is consistent with the observed slowdown in TOC reduction after the initial rapid degradation phase. Conversely, the EC values increased with time, reflecting the diminishing reactivity of the electro-Fenton process as the concentration of easily oxidizable organics decreased. The higher EC values at later stages of the process are attributed to the energy required to oxidize the remaining recalcitrant intermediates, which often involve complex molecular structures or require stronger oxidative conditions [54].

### 3.2. Electron Paramagnetic Resonance (EPR) Spectroscopy

By EPR spectroscopy, radical species generated during the electrochemical degradation of paracetamol (N-acetyl-para-aminophenol) were carefully analyzed and characterized. The spin-trapping molecule DMPO/TMPO was employed to react with radicals generated during the electrochemical reaction, forming a stable adduct radical.

Two different adduct radicals were identified I around $g = 2.0047$, in the EPR spectra, with a lifetime decay of a few minutes. As the electrochemical reaction progresses during electro-Fenton degradation, the concentration of radical species increases proportionally with the reaction time, underscoring the dependence of radical formation with the redox process of paracetamol and $H_2O_2$ [55].

Fig. 3 illustrates the chemical reaction between the spin-trapping molecule and the radicals generated in the electrochemical system. As the reaction progresses, the increase in radical concentration indicates the dynamic nature of radical formation, which is directly linked to the ongoing redox processes at the electrode surface.

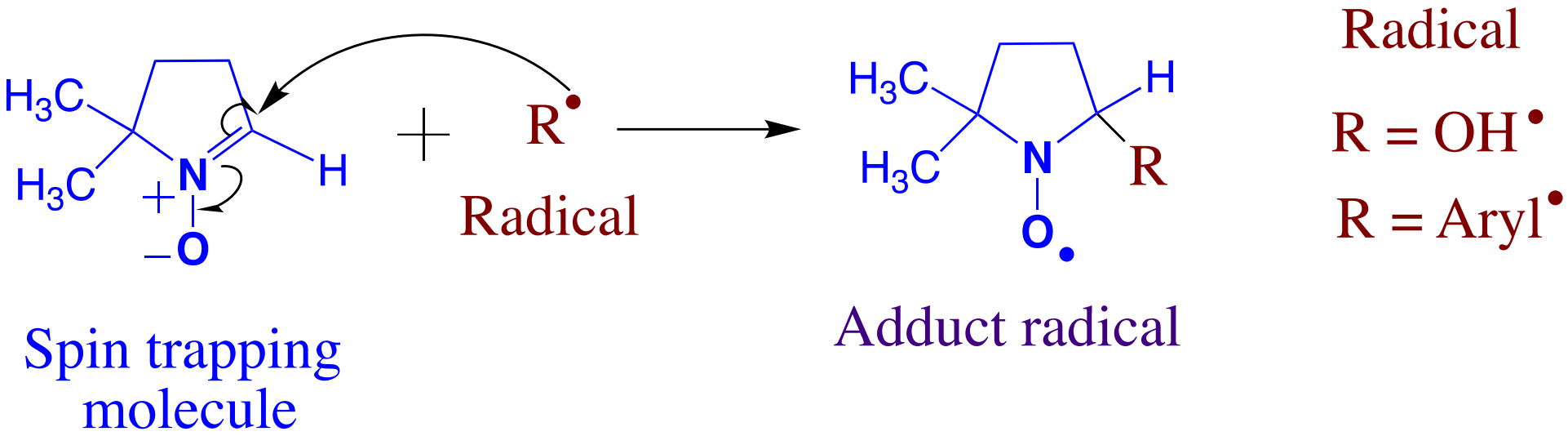

**Figure 3.** Chemical reaction between spin trapping molecule (DMPO) and radical species generated during the electrochemical reaction.

Fig. 4 shows the EPR signal of the radical adducts generated during the electrochemical reaction of paracetamol and $H_2O_2$ using a platinum (Pt) anode. The observed EPR signal was successfully simulated (Fig. 5) for hydroxyl radical using coupling constants of $a_N$ = 14.9 G and $a_H$ = 14.9 G.

The radical assignments was performed by a direct comparison between the experimental hyperfine coupling constants obtained in this work and those previously reported in the literature for other chemical derivatives of aryl radicals and hydroxyl adducts, which are in excellent agreement with the well-established values reported by Buettner [56].

The EPR spectra simulated for the spintrap–aryl radical adducts displayed coupling constants of $a_N$ = 16.0 G and $a_H$ = 24.0 G, values that closely match those reported for aryl-type radicals generated from photolyzed Organo-Boron compounds or ion-pair complexes [57,58], as also observed in arenediazonium compounds and aromatic intermediates of phenolic oxidation processes identified by Reszka and coworkers [60].

The close correspondence between our coupling constant parameters and those found in the literature provides strong evidence supporting our assignment of the secondary radical species as aryl-type intermediates derived from the hydroxylation and subsequent oxidation of the paracetamol aromatic ring.

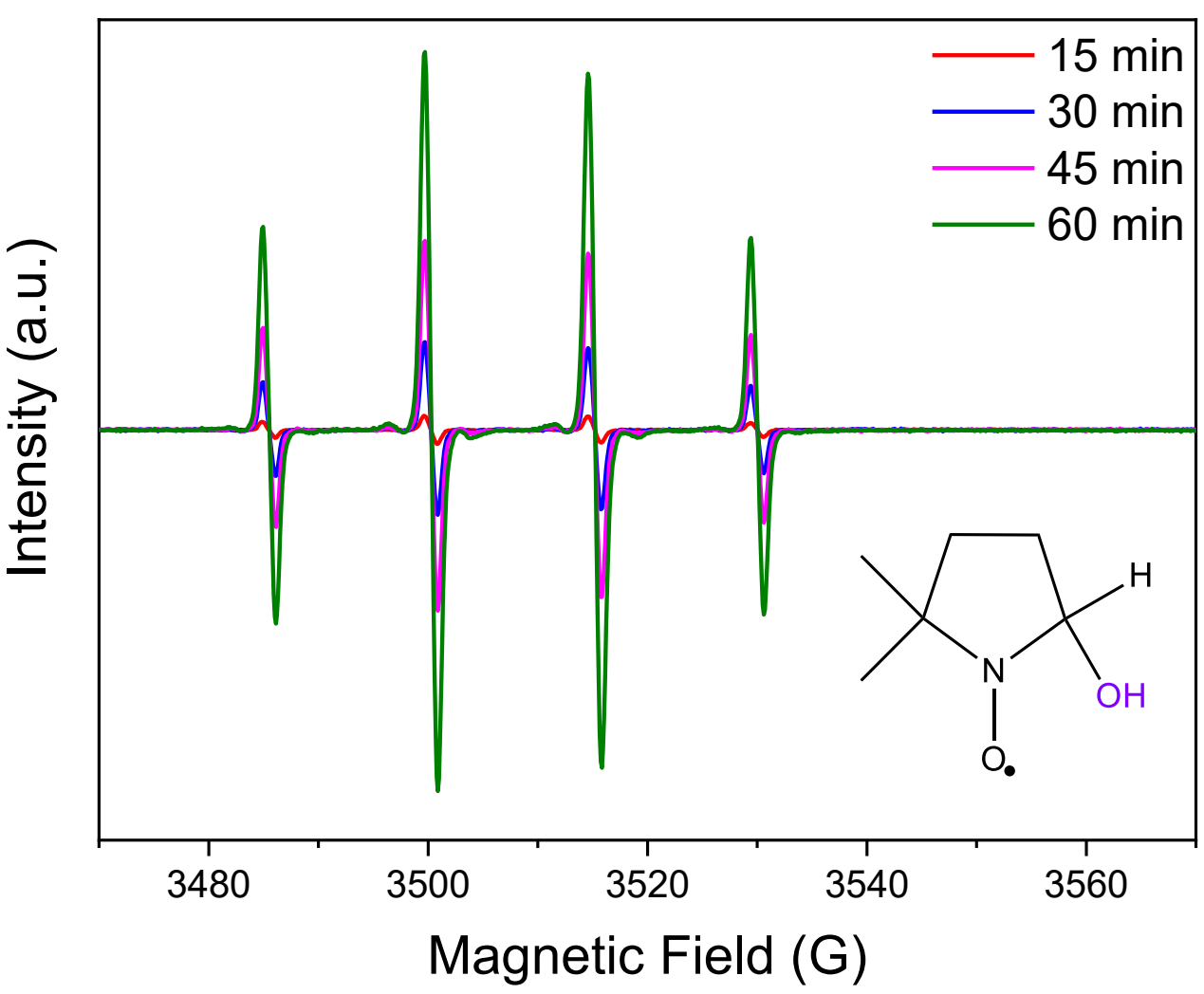

**Figure 4.** EPR spectra were recorded at different times of the electrochemical reaction. Chemical system: aqueous solution of 0.1 mol $L^{-1}$ $K_2SO_4$, 3.3 × $10^{-4}$ mol $L^{-1}$ N-acetyl-para-aminophenol and 0.5 × $10^{-3}$ mol $L^{-1}$ $Fe_2SO_4$. pH = 3.0; Platinum electrode was used as anode electrode; DMPO was used as a spin trapping molecule, and the microwave power intensity was fixed at 5 mW for all EPR measurements. Other experimental parameters, such as magnetic field sweep (100 G), modulation amplitude (1.0 G), conversion time (40.96 ms), and receiver gain (1.0 × $10^5$) were used in all measurements.

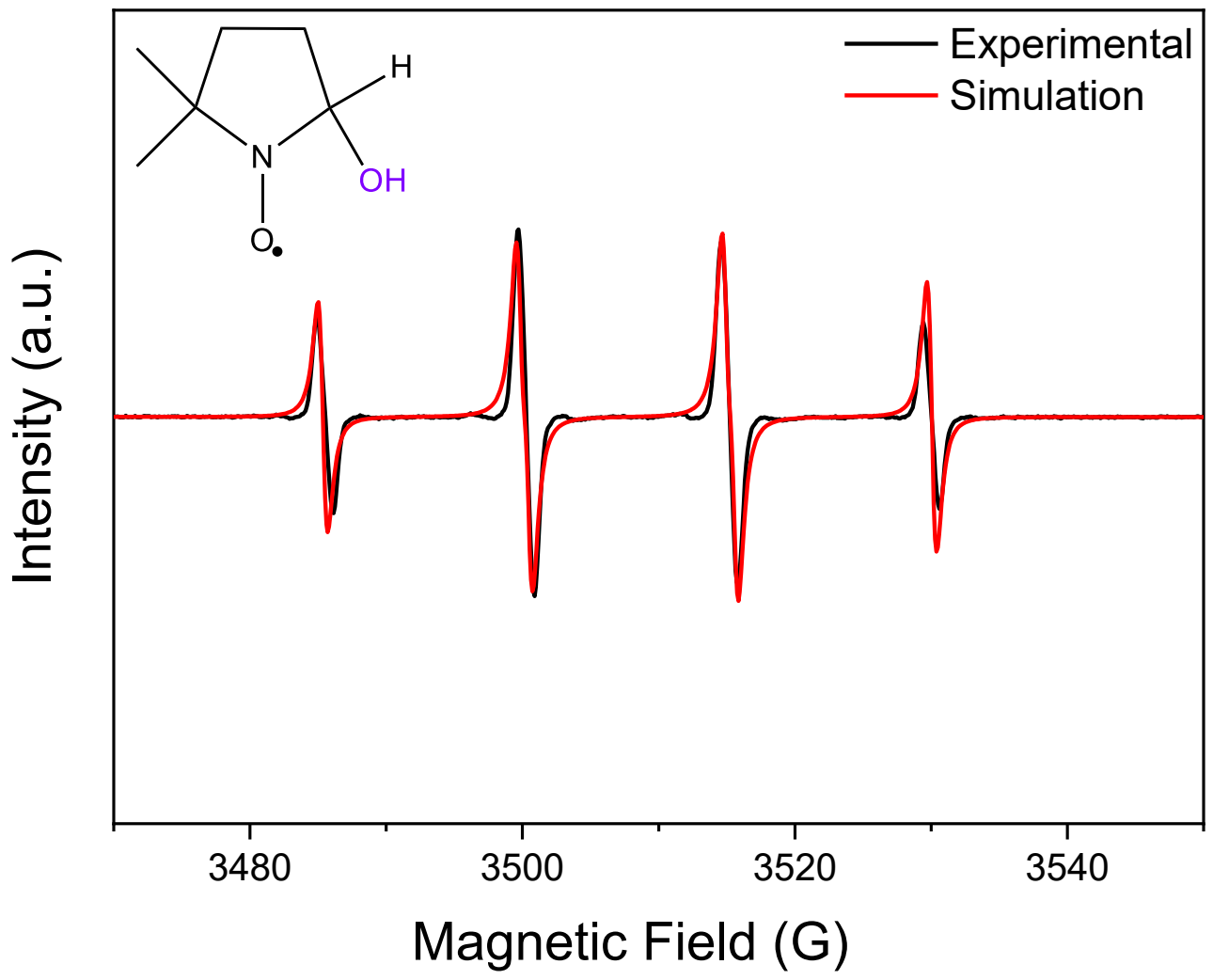

**Figure 5.** Experimental and simulated EPR spectra of the electrochemical system at 60 min of reaction. Chemical system: 0.1 mol $L^{-1}$ $K_2SO_4$, 3.3 × $10^{-4}$ mol $L^{-1}$ N-acetyl-para-aminophenol and 0.5 × $10^{-3}$ mol $L^{-1}$ $Fe_2SO_4$. pH = 3.0; platinum electrode was used as anode electrode. EPR Spectrum was simulated with $a_N$ = 14.9 and $a_H$ = 14.9 as constant parameters. Other experimental parameters, such as magnetic field sweep (100 G), modulation amplitude (1.0 G), conversion time (40.96 ms), and receiver gain (1.0 × $10^5$)

were used in all measurements and/or simulations. DMPO was used as a spin-trapping molecule.

In the early stages of the reaction (around 10 seconds), a second radical component was observed in the EPR spectra, with a small contribution (~2%) and g-values of 1.990. This suggests the presence of an organic radical species, likely an intermediate formed through the interaction of $^{\bullet}OH$ with the paracetamol structure. However, the characterization of this second radical was hindered by background noise, which prevented a definitive identification. This observation raises important questions about the potential formation of additional organic radical intermediates that may be involved in the complex degradation pathways.

When the BDD anode was used, the spin-trapping molecule was changed to TMPO, which allowed for the detection of two distinct radical species in the EPR spectra (Fig. 6). Both adduct-radical signals were simulated with the following hyperfine constant parameters: (i) $a_N = 16.2$ G, $a_H = 2.3$ G, with 75% contribution, and (ii) $a_N = 16.0$ G, $a_H = 3.2$ G, with 26% contribution. The primary species (74%) was attributed to $^{\bullet}OH$, while the second species (26%) was attributed to an aromatic ring radical adduct formed due to the attack of $^{\bullet}OH$ on the paracetamol molecule [57]. This finding is significant because it indicates that the BDD anode facilitates the formation of hydroxyl radicals and the generation of aromatic radicals, suggesting a more complex degradation pathway.

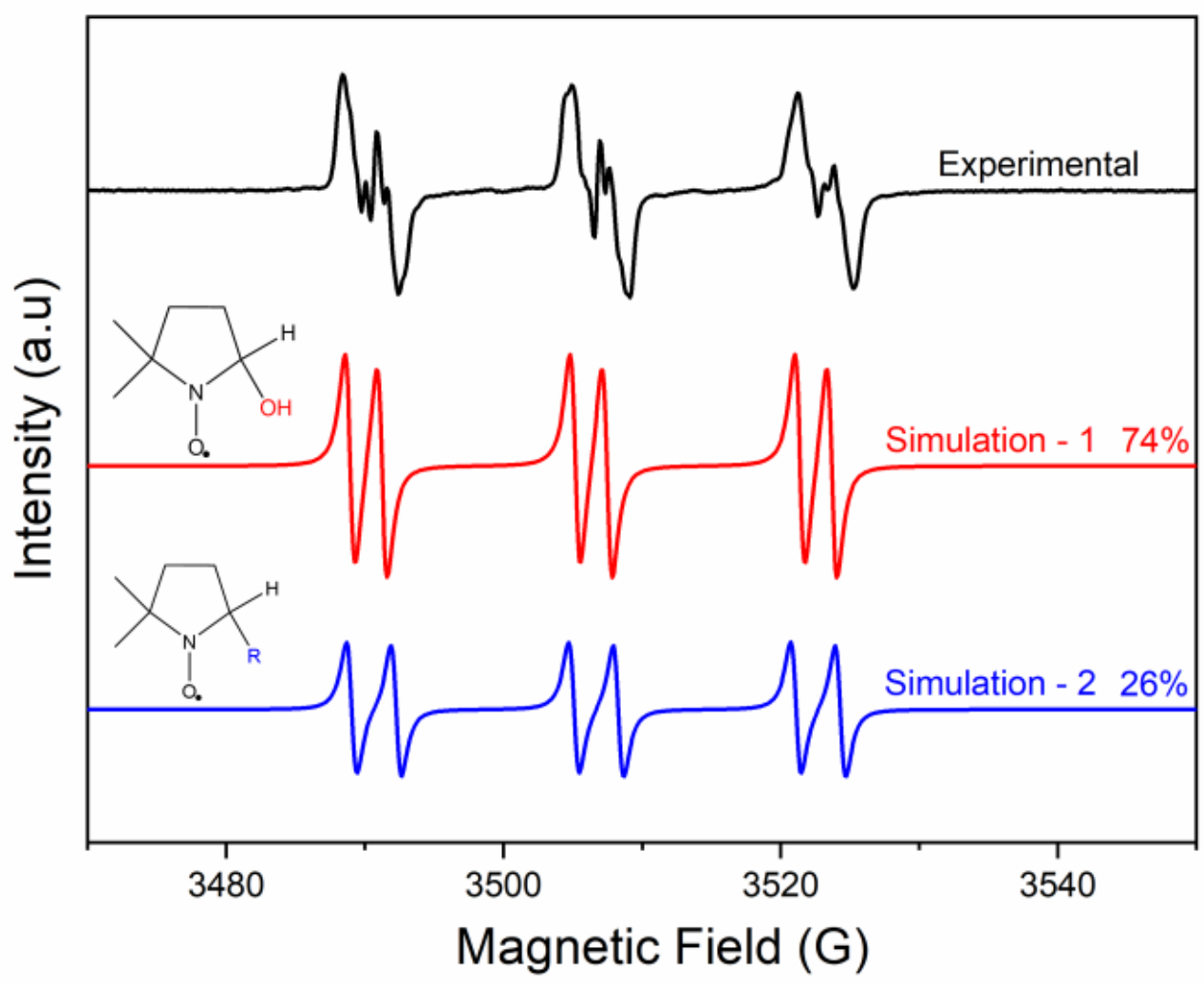

**Figure 6.** Experimental and simulated EPR spectra showing two adduct-radical components at 90 min of electrochemical reaction. Chemical system: 0.1 mol $L^{-1}$ $K_2SO_4$, $3.3 \times 10^{-4}$ mol $L^{-1}$ N-acetyl-para-aminophenol and $0.5 \times 10^{-3}$ mol $L^{-1}$ $Fe_2SO_4$. pH = 3.0;

Pt electrode was used as anode electrode. Simulation-1 signal was simulated with $a_N$ = 16.2, $a_H$ = 2.3, and 74% of the contribution percentage. Simulation-2 signal was simulated with $a_N$ = 16.0, $a_H$ = 3.2, and 26% of the contribution percentage. Other experimental parameters, such as magnetic field sweep (100 G), modulation amplitude (1.0 G), conversion time (40.96 ms), and receiver gain ($1.0 \times 10^5$) were used in all measurements and/or simulations.

The EPR analysis using TMPO further revealed the presence of two distinct stable radical species (Fig. 7), highlighting the involvement of different radical pathways during the electro-Fenton degradation of paracetamol. The hyperfine coupling constants for these species were determined to be $a_N$ = 15.5 G, $a_H$ = 16.6 G (65%) for the first species, and $a_N$ = 16.0 G, $a_H$ = 24.0 G (35%) for the second species (Fig. 8). The first species is attributed to $^{\bullet}OH$. In contrast, the second is linked to aryl radicals formed due to the degradation of paracetamol [58].

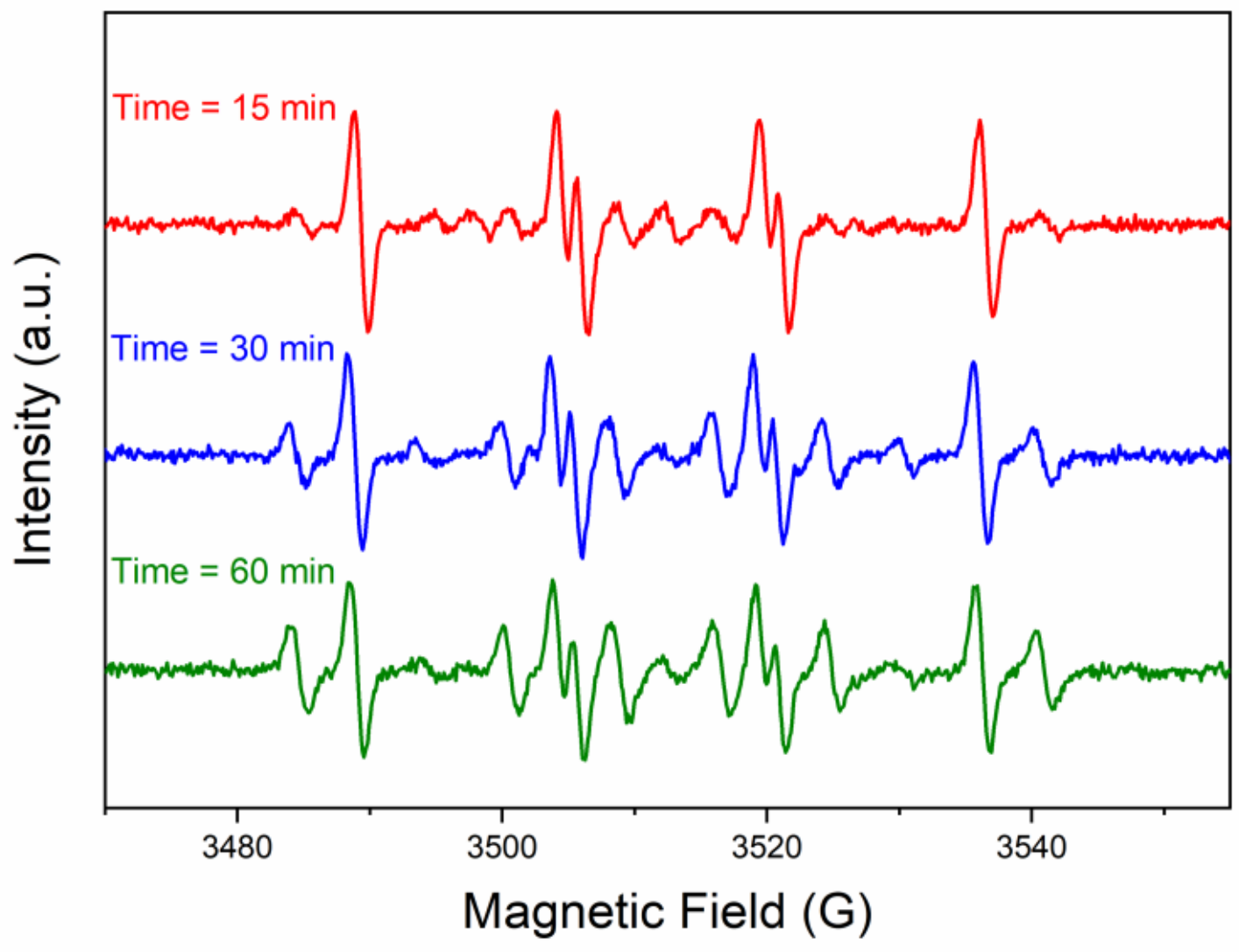

**Figure 7.** EPR spectra were recorded at different times of electrochemical reaction. Chemical system: aqueous solution of 0.1 mol $L^{-1}$ $K_2SO_4$, $3.3 \times 10^{-4}$ mol $L^{-1}$ N-acetyl-para-aminophenol and $0.5 \times 10^{-3}$ mol $L^{-1}$ $Fe_2SO_4$. BDD anode, pH was fixed at 3.0, TMPO was used as spin trapping molecule, and the microwave power intensity was fixed at 5 mW for all EPR measurements. Other experimental parameters, such as magnetic field sweep (100 G), modulation amplitude (1.0 G), conversion time (40.96 ms), and receiver gain ($1.0 \times 10^5$) were used in all measurements.

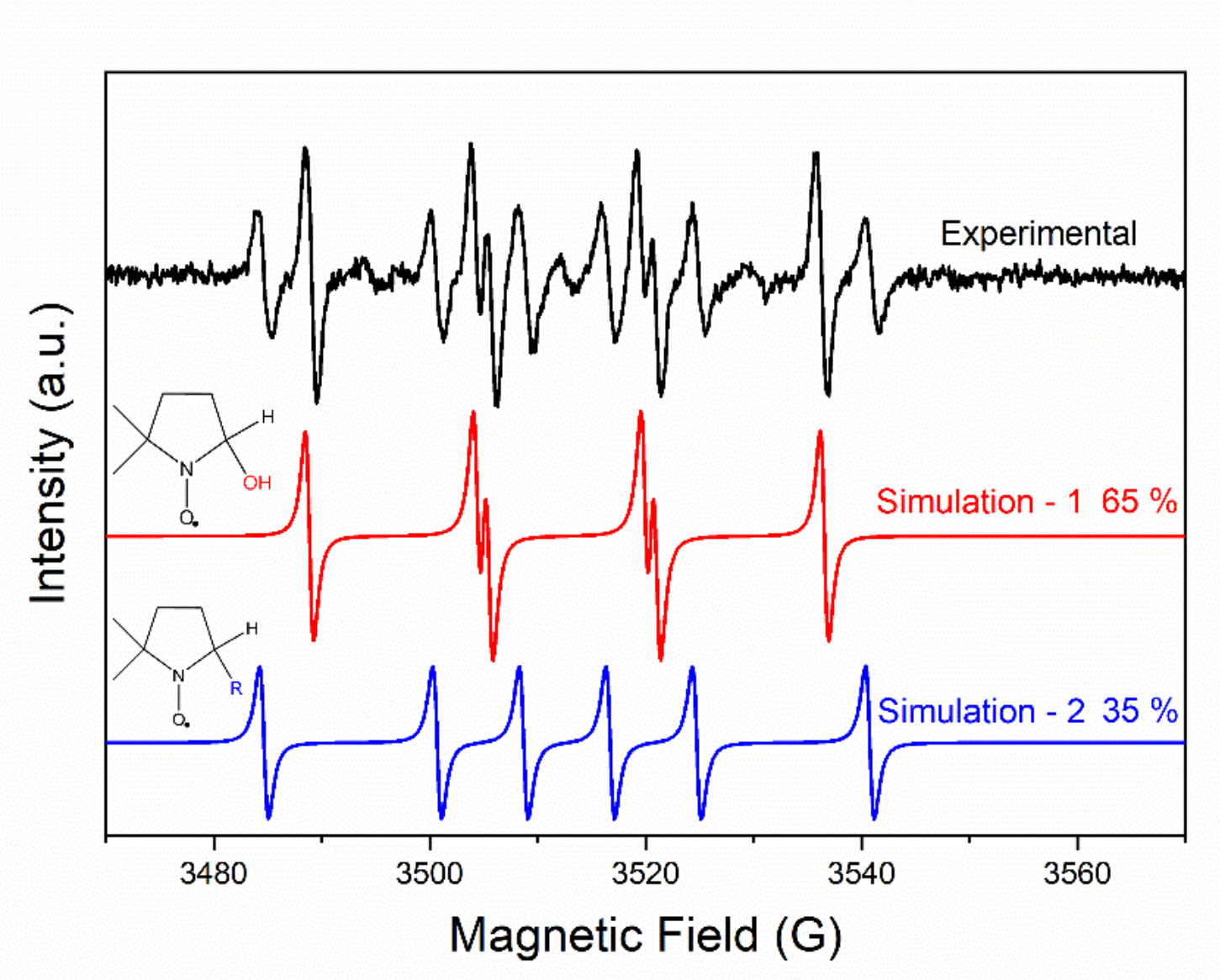

**Figure 8.** Experimental and simulated EPR spectra at 60 min of electrochemical reaction. Chemical system: 0.1 mol $L^{-1}$ $K_2SO_4$, 3.3 × $10^{-4}$ mol $L^{-1}$ N-acetyl-para-aminophenol and 0.5 × $10^{-3}$ mol $L^{-1}$ $Fe_2SO_4$. BDD anode pH was fixed at 3.0. EPR spectrum was simulated with $a_N$ = 15.5 and $a_H$ = 16.6 for simulation-1 and $a_N$ = 16.0 and $a_H$ = 24.0 for simulation-2. TMPO was used as a spin trap (1×$10^{-4}$ mol $L^{-1}$). Microwave power (5 mW), magnetic field sweep (100 G), modulation amplitude (1.0 G), conversion time (40.96 ms), and receiver gain (1.0 × $10^5$) were used in all measurements and/or simulations.

A key observation from the comparison between the Pt and BDD anodes is the shift in radical distribution. With the Pt anode, the contribution of $^•$OH was 74%, with only 26% attributed to aryl radicals. However, with the BDD anode, the relative contribution of aryl radicals increased to 35%, with a corresponding decrease in $^•$OH production (65%). This shift suggests that the BDD anode promotes a more complex degradation pathway, where aryl radicals—likely generated through electron transfer mechanisms at the electrode surface—play a more prominent role in the breakdown of paracetamol. The increased formation of aryl radicals highlights the enhanced oxidative power of the BDD anode, which likely promotes direct oxidation of the paracetamol molecule in addition to the classical $^•$OH-mediated oxidation.

The higher formation of aryl radicals when using the BDD anode, along with the overall increased radical production, emphasizes the superior performance of the BDD anode in electrochemical degradation systems. This enhanced radical formation is attributed to the unique properties of the BDD anode, which enables the accumulation of reactive oxygen species (ROS) at the surface, leading to more efficient radical generation. In particular, the BDD anode's high overpotential for water oxidation facilitates the generation of hydroxyl and organic radicals, thereby improving the overall efficiency of the electro-Fenton degradation process.

In conclusion, the EPR results from this study confirm the formation of $^{\bullet}OH$ and aryl radicals during the electrochemical degradation of paracetamol and reveal essential differences in radical distribution depending on the electrode material. With its enhanced radical generation capabilities, the BDD anode promotes a more effective degradation of paracetamol through multiple radical pathways, suggesting its potential for more efficient treatment of organic pollutants in electrochemical advanced oxidation processes.

### 3.3. Comparison with the literature

When comparing the $NaNbO_3$@$CeO_2$ cathode coupled with a BDD anode for paracetamol degradation via electro-Fenton with other previously published works, the present system demonstrates a highly competitive, if not superior, performance. For instance, while systems utilizing heteroatom-doped graphene materials achieved 100% degradation in 240 minutes and 52% mineralization of paracetamol in 360 minutes [61], and those based on porous cathode from carbon felt by thermal treatment achieved around 80% mineralization in 420 minutes [62], our work achieves a significantly higher mineralization of 81.6% in a remarkably 60-minute timeframe and 100% degradation in 15 minutes. However, the fundamental novelty of this work extends beyond mere performance metrics. While previous studies often rely on inferred mechanisms or monitor only the decay of the parent compound, the principal advancement here is the direct and quantitative mechanistic insight provided by in situ EPR spectroscopy. This approach uniquely allowed us to move beyond the common reporting of 'hydroxyl radical activity' to definitively quantify the specific contributions of hydroxyl versus aryl radicals (65% vs. 35%), directly linking the BDD anode to a more effective degradation pathway. Furthermore, the synergistic design of the $NaNbO_3$@$CeO_2$ heterostructure, which leverages the high oxygen mobility of $CeO_2$ and the stable catalytic properties of $NaNbO_3$, provides a more robust and active surface for $H_2O_2$ generation compared to single-metal oxide or standard carbon-based cathodes reported in prior studies [63-67]. Therefore, this work provides a significant advance by not only reporting a highly efficient system but also by establishing a definitive, quantitatively supported structure-activity relationship that bridges material design, radical generation mechanisms, and overall process efficiency, a crucial step for the rational design of next-generation electrochemical water treatment technologies.

## 4. Conclusions

This study successfully demonstrates the electrochemical degradation of paracetamol using a novel gas diffusion electrode (GDE) modified with $NaNbO_3$ nanocubes and $CeO_2$ nanorods, comparing Platinum (Pt) and boron-doped diamond (BDD) anodes. This work fundamentally advances the field by moving beyond simple performance comparisons to provide a quantitative, mechanistic explanation for the superior efficacy of BDD anodes. Through direct Electron Paramagnetic Resonance (EPR) spectroscopy, we uniquely quantified the radical species involved, revealing that the Pt system (74% $^{\bullet}OH$, 26% aryl radicals) relies predominantly on indirect oxidation, while the BDD system (65% $^{\bullet}OH$, 35% aryl radicals) leverages a more effective synergy between indirect hydroxyl radical attack and direct electron transfer at the anode surface. This mechanistic insight, directly correlating the anode material with specific radical generation profiles, explains the drastically faster degradation (15 min vs. 45 min) and higher mineralization (81.6% vs. 67.8%) achieved with BDD. Consequently, our primary contribution is the establishment of a definitive structure-activity relationship that bridges material selection and degradation mechanism. We conclusively show that the BDD anode, when paired with the $NaNbO_3$@$CeO_2$-modified GDE, is not merely a better performer but operates on a fundamentally more efficient principle. This combination represents a promising strategy for the rapid and effective removal of organic pollutants in electrochemical water treatment applications.

## CRediT authorship contribution statement

**Caio Machado Fernandes:** Investigation, Validation, Data curation, Writing – original draft. **João Paulo C. Moura:** Validation, Writing – review & editing. **Aline B. Trench:** Validation, Writing – review & editing. **Rafael Sotana:** Investigation, Writing – original draft. **Ana Maria P. Neto:** Writing – review & editing, **Willy G. Santos:** Conceptualization, Investigation, Validation, Data curation, Writing – original draft. **Mauro C. Santos:** Conceptualization, Writing – review & editing, Supervision.

## Data availability

The raw/processed data required to reproduce these findings cannot be shared for legal or ethical reasons.

**Declaration of Competing Interest**

The authors declare that they have no known competing financial interests or personal relationships that could have appeared to influence the work reported in this paper.

**Acknowledgments**

The authors would like to thank Fundação de Amparo à Pesquisa do Estado de São Paulo (FAPESP, #2021/05364-7, #2021/14394-7, #2022/10484-4, #2022/12895-1, and #2022/15252-4) for the financial support. The authors are also grateful for Coordenação de Aperfeiçoamento de Pessoal de Nível Superior (CAPES) and Conselho Nacional de Desenvolvimento Científico e Tecnológico (CNPq) (#303943/2021-1, #308663/2023–3, #402609/2023–9) for their support.

**References:**

[1] V.S. Pinheiro, E.C. Paz, L.R. Aveiro, L.S. Parreira, F.M. Souza, P.H.C. Camargo, M.C. Santos, Mineralization of paracetamol using a gas diffusion electrode modified with ceria high aspect ratio nanostructures, Electrochimica Acta 295 (2019) 39.
[2] E. Nagles, M. Ceroni, J.J. Hurtado-Murillo, J.J. Hurtado, Electrochemical determination of paracetamol in a pharmaceutical dose by adsorptive voltammetry with a carbon paste/La2O3 microcomposite, Analytical Methods 12 (2020) 2608.
[3] S. Periyasamy, M. Muthuchamy, Electrochemical oxidation of paracetamol in water by graphite anode: Effect of pH, electrolyte concentration and current density, Journal of Environmental Chemical Engineering 6 (2018) 7358.
[4] M. Pacheco-Álvarez, R. Picos Benítez, O.M. Rodríguez-Narváez, E. Brillas, J.M. Peralta-Hernández, A critical review on paracetamol removal from different aqueous matrices by Fenton and Fenton-based processes, and their combined methods, Chemosphere 303 (2022) 134883.
[5] C. Machado Fernandes, G.A. Cerron-Calle, E. Brillas, M.C. Santos, S. Garcia-Segura, Optimization of fipronil removal via electro-Fenton using a carbon cloth air-diffusion electrode, Separation and Purification Technology 367 (2025) 132955.
[6] C. Machado Fernandes, E. Brillas, M.C. Santos, S. Garcia-Segura, Electro-Fenton treatment of benzophenone-4 solutions: A sustainable approach for its removal using an air-diffusion cathode, Process Safety and Environmental Protection 199 (2025) 107342.
[7] X. Sun, W. Du, Y. Zu, Spinel sulfide with hierarchical porous structure for efficient degradation of metronidazole in electro-Fenton process, Colloids and Surfaces A: Physicochemical and Engineering Aspects 722 (2025) 137257.
[8] A.B. Trench, C.M. Fernandes, J.P.C. Moura, L.E.B. Lucchetti, T.S. Lima, V.S. Antonin, J.M. de Almeida, P. Autreto, I. Robles, A.J. Motheo, M.R.V. Lanza, M.C. Santos, Hydrogen peroxide electrogeneration from O(2) electroreduction: A review focusing on carbon electrocatalysts and environmental applications, Chemosphere 352 (2024) 141456.

[9] I. Lozano, P. Cervantes-Aviles, A. Keller, C.L. Aguilar, Removal of pharmaceuticals and personal care products from wastewater via anodic oxidation and electro-Fenton processes: current status and needs regarding their application, Water Science and Technology 88 (2023) 1143.
[10] F. Deng, H. Olvera-Vargas, M. Zhou, S. Qiu, I. Sirés, E. Brillas, Critical Review on the Mechanisms of Fe2+ Regeneration in the Electro-Fenton Process: Fundamentals and Boosting Strategies, Chemical Reviews 123 (2023) 4635.
[11] J. Hu, B. Wu, L. Chen, C. Song, H. Yang, F. Long, J. Sun, R. Chi, Influences of CeO2 morphology on enhanced performance of electro-Fenton for wastewater treatment, Journal of Rare Earths 40 (2022) 1870.
[12] E. Brillas, I. Sirés, M.A. Oturan, Electro-Fenton Process and Related Electrochemical Technologies Based on Fenton's Reaction Chemistry, Chemical Reviews 109 (2009) 6570.
[13] J.P.C. Moura, L.E.B. Lucchetti, C.M. Fernandes, A.B. Trench, C.N. Lange, B.L. Batista, J.M. Almeida, M.C. Santos, Experimental and theoretical studies of WO3/Vulcan XC-72 electrocatalyst enhanced H2O2 yield ORR performed in acid and alkaline medium, Journal of Environmental Chemical Engineering 12 (2024) 113182.
[14] X. Cao, D. Jiang, M. Huang, J. Pan, J. Lin, W. Chan, Iron oxide nanoparticles wrapped in graphene aerogel composite: Fabrication and application in electro-fenton at a Wide pH, Colloids and Surfaces A: Physicochemical and Engineering Aspects 587 (2020) 124269.
[15] Y. Sun, S. Tu, Y. Li, X. Sui, S. Geng, H. Wang, X. Duan, L. Chang, Degradation mechanism of methylene blue by heterogenous electro-Fenton with CeO2/rGO composite cathode, Colloids and Surfaces A: Physicochemical and Engineering Aspects 690 (2024) 133861.
[16] S. Tu, Z. Ning, X. Duan, X. Zhao, L. Chang, Efficient electrochemical hydrogen peroxide generation using TiO2/rGO catalyst and its application in electro-Fenton degradation of methyl orange, Colloids and Surfaces A: Physicochemical and Engineering Aspects 651 (2022) 129657.
[17] H. Huang, X. Zou, R. Ji, J. Zhang, Z. Yuan, M. Zhao, H. Zhang, J. Geng, J. Li, Inducing three-phase interface to enhance hydroxyl radical production via green atomic H*-mediated electro-Fenton process for highly-efficient tetracycline degradation, Colloids and Surfaces A: Physicochemical and Engineering Aspects 698 (2024) 134577.
[18] M.H.M.T. Assumpção, A. Moraes, R.F.B. De Souza, I. Gaubeur, R.T.S. Oliveira, V.S. Antonin, G.R.P. Malpass, R.S. Rocha, M.L. Calegaro, M.R.V. Lanza, M.C. Santos, Low content cerium oxide nanoparticles on carbon for hydrogen peroxide electrosynthesis, Applied Catalysis A: General 411-412 (2012) 1-6.
[19] M.H.M.T. Assumpção, A. Moraes, R.F.B. De Souza, M.L. Calegaro, M.R.V. Lanza, E.R. Leite, M.A.L. Cordeiro, P. Hammer, M.C. Santos, Influence of the preparation method and the support on H2O2 electrogeneration using cerium oxide nanoparticles, Electrochimica Acta 111 (2013) 339-343.
[20] C. Machado Fernandes, A.O. Santos, V.S. Antonin, J.P.C. Moura, A.B. Trench, O.C. Alves, Y. Xing, J.C.M. Silva, M.C. Santos, Magnetic field-enhanced oxygen reduction reaction for electrochemical hydrogen peroxide production with different cerium oxide nanostructures, Chemical Engineering Journal 488 (2024) 150947.

[21] C. Machado Fernandes, J.P.C. Moura, A.B. Trench, O.C. Alves, Y. Xing, M.R.V. Lanza, J.C.M. Silva, M.C. Santos, Magnetic field-enhanced two-electron oxygen reduction reaction using CeMnCo nanoparticles supported on different carbonaceous matrices, Materials Today Nano 28 (2024) 100524.
[22] A.B. Trench, J. Paulo C. Moura, V.S. Antonin, C. Machado Fernandes, L. Liu, M.C. Santos, Improvement of H2O2 electrogeneration using a Vulcan XC72 carbon-based electrocatalyst modified with Ce-doped Nb2O5, Advanced Powder Technology 35 (2024) 104404.
[23] W. Sun, Y. Zhou, H. Zou, G. Liu, Synergistic spatial confinement and electron penetration for ultrafast Fenton decontamination of micro pollutants and nanoplastics, Journal of Hazardous Materials 495 (2025) 139150.
[24] W. Sun, G. Liu, H. Zou, S. Wang, X. Duan, Site-designed dual-active-center catalysts for co-catalysis in advanced oxidation processes, npj Materials Sustainability 3 (2025) 2.
[25] J. Teng, Y. Zhang, P. Li, G. Liu, Hydroxyl radical in anodic oxidation-based electrochemical advanced oxidation processes: Critical understanding and a new perspective, Sustainable Horizons 16 (2025) 100156.
[26] V.S. Antonin, L.E.B. Lucchetti, F.M. Souza, V.S. Pinheiro, J.P.C. Moura, A.B. Trench, J.M. de Almeida, P.A.S. Autreto, M.R.V. Lanza, M.C. Santos, Sodium niobate microcubes decorated with ceria nanorods for hydrogen peroxide electrogeneration: An experimental and theoretical study, Journal of Alloys and Compounds 965 (2023) 171363.
[27] V.S. Antonin, F.M. Souza, V.S. Pinheiro, J.P.C. Moura, A.B. Trench, C.M. Fernandes, M.R.V. Lanza, M.C. Santos, Electrocatalytic hydrogen peroxide generation using WO3 nanoparticle-decorated sodium niobate microcubes, Journal of Electroanalytical Chemistry 959 (2024) 118190.
[28] C.-C. Su, A.-T. Chang, L.M. Bellotindos, M.-C. Lu, Degradation of acetaminophen by Fenton and electro-Fenton processes in aerator reactor, Separation and Purification Technology 99 (2012) 8.
[29] T.X.H. Le, C. Charmette, M. Bechelany, M. Cretin, Facile Preparation of Porous Carbon Cathode to Eliminate Paracetamol in Aqueous Medium Using Electro-Fenton System, Electrochimica Acta 188 (2016) 378.
[30] K. Shi, Y. Wang, A. Xu, H. Zhu, L. Gu, X. Liu, J. Shen, W. Han, K. Wei, Integrated electro-Fenton system based on embedded U-tube GDE for efficient degradation of ibuprofen, Chemosphere 311 (2023) 137196.
[31] A. Cruz-Rizo, S. Gutiérrez-Granados, R. Salazar, J.M. Peralta-Hernández, Application of electro-Fenton/BDD process for treating tannery wastewaters with industrial dyes, Separation and Purification Technology 172 (2017) 296.
[32] A. Sennaoui, S. Alahiane, F. Sakr, M. Tamimi, E.H. Ait Addi, M. Hamdani, A. Assabbane, Comparative degradation of benzoic acid and its hydroxylated derivatives by electro-Fenton technology using BDD/carbon-felt cells, Journal of Environmental Chemical Engineering 7 (2019) 103033.
[33] M.A. Oturan, Outstanding performances of the BDD film anode in electro-Fenton process: Applications and comparative performance, Current Opinion in Solid State and Materials Science 25 (2021) 100925.
[34] M. Mbaye, P.A. Diaw, O.M.A. Mbaye, N. Oturan, M.D. Gaye Seye, C. Trellu, A. Coly, A. Tine, J.-J. Aaron, M.A. Oturan, Rapid removal of fungicide thiram in aqueous

medium by electro-Fenton process with Pt and BDD anodes, Separation and Purification Technology 281 (2022) 119837.
[35] H. Chen, G. Yoshida, F.J. Andriamanohiarisoamanana, I. Ihara, Electro-oxidation combined with electro-Fenton for decolorization of caramel colorant aqueous solution using BDD electrodes, Journal of Water Process Engineering 47 (2022) 102672.
[36] O. García, E. Isarain-Chávez, S. Garcia-Segura, E. Brillas, J.M. Peralta-Hernández, Degradation of 2,4-Dichlorophenoxyacetic Acid by Electro-oxidation and Electro-Fenton/BDD Processes Using a Pre-pilot Plant, Electrocatalysis 4 (2013) 224.
[37] K. Cruz-González, O. Torres-Lopez, A.M. García-León, E. Brillas, A. Hernández-Ramírez, J.M. Peralta-Hernández, Optimization of electro-Fenton/BDD process for decolorization of a model azo dye wastewater by means of response surface methodology, Desalination 286 (2012) 63.
[38] K. Zuo, S. Garcia-Segura, G.A. Cerrón-Calle, F.-Y. Chen, X. Tian, X. Wang, X. Huang, H. Wang, P.J.J. Alvarez, J. Lou, M. Elimelech, Q. Li, Electrified water treatment: fundamentals and roles of electrode materials, Nature Reviews Materials 8 (2023) 472.
[39] G. Song, X. Du, Y. Zheng, P. Su, Y. Tang, M. Zhou, A novel electro-Fenton process coupled with sulfite: Enhanced Fe3+ reduction and TOC removal, Journal of Hazardous Materials 422 (2022) 126888.
[40] H. Öztürk, S. Barışçı, O. Turkay, Paracetamol degradation and kinetics by advanced oxidation processes (AOPs): Electro-peroxone, ozonation, goethite catalyzed electro-fenton and electro-oxidation, Environmental Engineering Research 26 (2021) 180332.
[41] F. Ghanbari, A. Hassani, S. Wacławek, Z. Wang, G. Matyszczak, K.-Y.A. Lin, M. Dolatabadi, Insights into paracetamol degradation in aqueous solutions by ultrasound-assisted heterogeneous electro-Fenton process: Key operating parameters, mineralization and toxicity assessment, Separation and Purification Technology 266 (2021) 118533.
[42] O. Ganzenko, C. Trellu, N. Oturan, D. Huguenot, Y. Péchaud, E.D. van Hullebusch, M.A. Oturan, Electro-Fenton treatment of a complex pharmaceutical mixture: Mineralization efficiency and biodegradability enhancement, Chemosphere 253 (2020) 126659.
[43] H. Olvera-Vargas, J.-C. Rouch, C. Coetsier, M. Cretin, C. Causserand, Dynamic cross-flow electro-Fenton process coupled to anodic oxidation for wastewater treatment: Application to the degradation of acetaminophen, Separation and Purification Technology 203 (2018) 143-151.
[44] W. Yang, N. Oturan, J. Liang, M.A. Oturan, Synergistic mineralization of ofloxacin in electro-Fenton process with BDD anode: Reactivity and mechanism, Separation and Purification Technology 319 (2023) 124039.
[45] J. Cai, M. Zhou, Y. Pan, X. Lu, Degradation of 2,4-dichlorophenoxyacetic acid by anodic oxidation and electro-Fenton using BDD anode: Influencing factors and mechanism, Separation and Purification Technology 230 (2020) 115867.
[46] H. Olvera-Vargas, N. Gore-Datar, O. Garcia-Rodriguez, S. Mutnuri, O. Lefebvre, Electro-Fenton treatment of real pharmaceutical wastewater paired with a BDD

anode: Reaction mechanisms and respective contribution of homogeneous and heterogeneous OH, Chemical Engineering Journal 404 (2021) 126524.
[47] F.E. Titchou, H. Zazou, H. Afanga, J. El Gaayda, R. Ait Akbour, M. Hamdani, M.A. Oturan, Electro-Fenton process for the removal of Direct Red 23 using BDD anode in chloride and sulfate media, Journal of Electroanalytical Chemistry 897 (2021) 115560.
[48] S. Midassi, A. Bedoui, N. Bensalah, Efficient degradation of chloroquine drug by electro-Fenton oxidation: Effects of operating conditions and degradation mechanism, Chemosphere 260 (2020) 127558.
[49] M.F. García-Montoya, S. Gutiérrez-Granados, A. Alatorre-Ordaz, R. Galindo, R. Ornelas, J.M. Peralta-Hernández, Application of electrochemical/BDD process for the treatment wastewater effluents containing pharmaceutical compounds, Journal of Industrial and Engineering Chemistry 31 (2015) 238.
[50] N. Oturan, C.T. Aravindakumar, H. Olvera-Vargas, M.M. Sunil Paul, M.A. Oturan, Electro-Fenton oxidation of para-aminosalicylic acid: degradation kinetics and mineralization pathway using Pt/carbon-felt and BDD/carbon-felt cells, Environmental Science and Pollution Research 25 (2018) 20363.
[51] O. García, E. Isarain-Chávez, A. El-Ghenymy, E. Brillas, J.M. Peralta-Hernández, Degradation of 2,4-D herbicide in a recirculation flow plant with a Pt/air-diffusion and a BDD/BDD cell by electrochemical oxidation and electro-Fenton process, Journal of Electroanalytical Chemistry 728 (2014) 1.
[52] A. El-Ghenymy, R.M. Rodríguez, E. Brillas, N. Oturan, M.A. Oturan, Electro-Fenton degradation of the antibiotic sulfanilamide with Pt/carbon-felt and BDD/carbon-felt cells. Kinetics, reaction intermediates, and toxicity assessment, Environmental Science and Pollution Research 21 (2014) 8368.
[53] H. Zazou, N. Oturan, M. Sönmez-Çelebi, M. Hamdani, M.A. Oturan, Mineralization of chlorobenzene in aqueous medium by anodic oxidation and electro-Fenton processes using Pt or BDD anode and carbon felt cathode, Journal of Electroanalytical Chemistry 774 (2016) 22.
[54] A. Huang, D. Zhi, Y. Zhou. A novel modified Fe–Mn binary oxide graphite felt (FMBO-GF) cathode in a neutral electro-Fenton system for ciprofloxacin degradation, Environmental Pollutiton 286 (2021) 117310.
[55] Z. Wang, D. Ren, S. Shang, S. Zhang, X. Zhang, W. Chen, Novel synthesis of Cu-HAP/$SiO_2$@carbon nanocomposites as heterogeneous catalysts for Fenton-like oxidation of 2,4-DCP, Advanced Powder Technology 33 (2022) 103509.
[56] G.R. Buettner, Spin Trapping: ESR parameters of spin adducts 1474 1528V, Free Radical Biology and Medicine 3 (1987) 259.
[57] W.G. Santos, D.S. Budkina, S.H. Santagneli, A.N. Tarnovsky, J. Zukerman-Schpector, S.J.L. Ribeiro, Ion-Pair Complexes of Pyrylium and Tetraarylborate as New Host–Guest Dyes: Photoinduced Electron Transfer Promoting Radical Polymerization, The Journal of Physical Chemistry A 123 (2019) 7374.
S.B. Hammouda, C. Salazar, F. Zhao, D.L. Ramasamy, E. Laklova, S. Iftekhar, I. Babu, M. Sillanpää, Efficient heterogeneous electro -Fenton incineration of a contaminant of emergent concern-cotinine- in aqueous medium using the magnetic double perovskite oxide Sr2FeCuO6 as a highly stable catalayst: Degradation kinetics and oxidation products, Applied Catalysis B: Environmental 240 (2019) 201.

[58] W.G. Santos, D.S. Budkina, P.F.G.M. da Costa, D.R. Cardoso, A.N. Tarnovsky, M.D.E. Forbes, Inverse photochromism in viologen–tetraarylborate ion-pair complexes: optical write/microwave erase switching in polymer matrices, Materials Advances 3 (2022) 3862.
[60] K.J. Reszka, C.F. Chignell, One-electron reduction of arenediazonium compounds by physiological electron donors generates aryl radicals. An EPR and spin trapping investigation, Chemico-Biological Interactions 96 (1995) 223.
[61] N. Fernandez-Saez, D.E. Villela-Martinez, F. Carrasco-Marin, A.F. Perez-Cadenas, L.M. Pastrana-Martinez, Heteroatom-doped graphene aerogels and carbon-magnetite catalysts for the heterogeneous electro-Fenton degradation of acetaminophen in aqueous solution, Journal of Catalysis 378 (2019) 68.
[62] T.X.H. Le, C. Charmette, M. Bechelany, M. Cretin, Facile Preparation of Porous Carbon Cathode to Eliminate Paracetamol in Aqueous Medium Using Electro-Fenton System, Electrochimica Acta 188 (2016) 378.
[63] Q. Zhang, Y.-bo Yu, J.-ming Hong, Mechanism and efficiency research of P- and N-codoped graphene for enhanced paracetamol electrocatalytic degradation, Environmental Science and Pollution Research 29 (2022) 80281.
[64] T.X.H. Le, T.V. Nguyen, Z.A. Yacouba, L. Zoungrana, F. Avril, D.L. Nguyen, E. Petit, J. Mendret, V. Bonniol, M. Bechelany, S. Lacour, G. Lesage, M. Cretin, Correlation between degradation pathway and toxicity of acetaminophen and its by-products by using the electro-Fenton process in aqueous media, Chemosphere 172 (2017) 1.
[65] H. Öztürk, S. Barışçı, O. Turkay, Paracetamol degradation and kinetics by advanced oxidation processes (AOPs): Electro-peroxone, ozonation, goethite catalyzed electro-fenton and electro-oxidation, Environmental Engineering Research 26 (2021) 180332.
[66] S. Herrera-Chávez, S. Gutierrez, M.A. Sandoval, E. Brillas, M. Pacheco-Álvarez, J.M. Peralta-Hernández, Sustainable Degradation of Acetaminophen by a Solar-Powered Electro-Fenton Process: A Green and Energy-Efficient Approach, Processes 13 (2025) 2633.
[67] I. Bavasso, C. Poggi, E. Petrucci, Enhanced degradation of paracetamol by combining UV with electrogenerated hydrogen peroxide and ozone, Journal of Water Process Engineering 34 (2020) 101102.